\documentclass[11pt]{article}
\usepackage{amsmath,amssymb,graphicx}
\usepackage{psfrag,subfigure,tabularx}

\def\leigh{Robert G. Leigh}
\def\uiucaddress{\small Department of Physics, University of Illinois, 1110 W. Green St.,
Urbana IL 61801-3080, U.S.A. }
\def\title{\Large { Fermions and the Sch/nrCFT Correspondence}}

\newcommand\Dsl{D\kern-7pt/\kern1pt}
\newcommand\ksl{k\kern-6pt/\kern1pt}
\newcommand\kapsl{\kappa\kern-6pt/\kern1pt}
\newcommand\kslsl{k\kern-6pt/\kern1pt\kern-5pt/}
\newcommand\vsl{v\kern-5pt/\kern1pt}
\newcommand\wsl{w\kern-6pt/\kern1pt}
\newcommand\xsl{x\kern-6pt/\kern1pt}
\newcommand\qsl{{\cal Q}}

\newcommand\donu[2]{#2}

\parskip=\baselineskip

\usepackage{epsfig}

\newcommand\cc[1]{#1^{^{\kern-6pt \circ}}\kern2pt}
\newcommand\comment[1]{}

\newcommand{\RR}{\mathbb{R}}

\newcommand{\h}{\hspace}

\newcommand{\pa}{\partial}

\newcommand{\beq}{\begin{equation}}
\newcommand{\eeq}{\end{equation}}
\newcommand{\beqn}{\begin{eqnarray}}
\newcommand{\eeqn}{\end{eqnarray}}

\def\dalemb#1#2{{\vbox{\hrule height .#2pt
\hbox{\vrule width.#2pt height#1pt \kern#1pt
\vrule width.#2pt}
\hrule height.#2pt}}}

\begin{document}
\renewcommand\author[1]{#1}

\begin{center}
\vspace{60pt}
\title
\end{center}
\vskip 2 cm
\centerline{{\bf
\author{\leigh} and
 \author{Nam Nguyen Hoang}
 }}
\vspace{.5cm}
\centerline{\it \uiucaddress}

\vspace{2cm}

\begin{abstract}
We consider the problem of Dirac fermions propagating on a spacetime of Schr\"odinger isometry and the associated boundary Euclidean two-point function of fermionic scaling operators of the holographic dual non-relativistic conformal theory. Paying careful attention to the representations of the Schr\"odinger algebra that appear in this problem, we show carefully how the on-shell action is constructed and how the boundary theory may be renormalized consistently by the inclusion of appropriate Galilean invariant boundary counterterms.
\end{abstract}

\pagebreak

\section{Introduction}

The gauge/gravity holographic correspondence has in recent years been applied to an ever widening array of interesting (presumably strongly coupled) quantum field theories. It is believed that the holographic dictionary can be applied to cases in which the asymptotic geometry of the dual spacetime has isometry that differs from that of AdS. Since this asymptotic symmetry is reflected in the space-time symmetry of the field theory, one can obtain non-Poincar\'e invariant theories. An interesting example was introduced by Son \cite{Son:2008ye}, in which the isometry is the conformal Schr\"odinger symmetry. This geometrizes both dilatations as well as a phase symmetry that is related to 'particle number'. An extension to finite temperature and density was described in \cite{Balasubramanian:2008dm,Herzog:2008wg,Maldacena:2008wh,Adams:2008wt}.

Although the problem of finite density is of most direct physical interest, the system poses some interesting problems even in the `vacuum' geometry that possesses the full Schr\"odinger isometry. We believe that these issues should be sorted out before the finite temperature and density cases can be fully appreciated, and in fact it is possible to do so with complete precision as exact analytic solutions exist, as they do in the relativistic  case.

The Schr\"odinger conformal group has been of interest for many years as the full symmetry of the free Schr\"odinger wave equation (as well as a few systems with certain interactions). The reader will find discussions in the literature of the representation theory in many papers, particularly in low spatial dimensional cases \cite{Levy-Leblond,Hagen:1972pd,Perroud:1977qh,Henkel:2003pu}.

The Euclidean correlation functions of scalar operators have been investigated by Refs. \cite{Son:2008ye,Fuertes:2009ex,Volovich:2009yh} and a variety of Lorentzian correlators were discussed extending the methods of Skenderis and van Rees \cite{Skenderis:2008dg,Skenderis:2008dh} in \cite{Leigh:2009eb}. In the latter paper, we showed in particular that the boundary renormalization program can be carried out for scalars. The boundary renormalization has not, to our knowledge, been completely sorted out particularly with respect to the metric, as the Fefferman-Graham expansion that governs the asymptotically AdS spacetimes is not available, and must be replaced by a more intricate structure.

In the present paper, we investigate fermionic operators in the vacuum $z=2$ Schr\"odinger geometry. The system is significantly different from the corresponding relativistic case, as might be expected from the above comments as well as the somewhat more rich representation theory. Nevertheless, we are able to show that a sensible Dirichlet problem exists for fermions. As in the relativistic case, the bulk on-shell action vanishes, and the on-shell action is determined entirely by boundary terms. These boundary terms are uniquely determined by the requirements of a sensible Dirichlet canonical structure and finiteness. In particular, we explain the structure of possible boundary terms (which are required to preserve the Galilean symmetry of the regulated boundary theory) and show that the on-shell action is finite with the inclusion of a finite number of local boundary counterterms.

\section{Background}

In standard coordinates, the metric of Schr\"odinger isometry may be
written \beq\label{metric} ds^2 = \frac{L^2}{z^2}\left(-
\frac{\beta^2}{z^2}dt^2 + 2dt d\xi + d\vec{x}^2 + dz^2\right) \eeq
The $\vec x$ are coordinates in $d$-dimensional space. Although we
are primarily interested in Euclidean correlator, we will be working
on the Lorentzian geometry, as the former can be easily obtained by
a Wick rotation. Rather than giving a conformal class as in the
relativistic (AdS) case, the quantity in parentheses is a metric of
the Bargman type, in which the coordinate $\xi$ is null. The Killing
field $N=\pa_\xi$ generates the central extension of the
Schr\"odinger algebra whose eigenvalue would be interpreted as
`mass' or `particle number' in a weakly coupled non-relativistic
particle theory. In the present context, fields propagating in the
bulk are to be taken to be equivariant with respect to $N$ \beq
N\Psi = in\psi \eeq and dual quasi-primary operators are labeled by
both conformal dimension and $n$.  The $\xi$-direction is taken to
be compact so that the spectrum of dual operators is discrete.

A convenient basis of orthonormalized ($\langle e^a,e^b\rangle=\eta^{ab}$) one-forms is
\beqn
e^0=\frac{L}{z}\left[ \frac{z}{\beta} d\xi-\frac{\beta}{z}dt\right],\ \ \
e^v=\frac{L}{\beta} d\xi,\ \ \
e^r=\frac{L}{z}dz,\ \ \
e^i=\frac{L}{z}dx^i
\eeqn
which are dual to the orthonormal basis vectors
\beq
e_0 = -\frac{z}{L}{\frac{z}{\beta}}\pa_t,\ \ \
e_v=  \frac{z}{L}\left[{\frac{z}{\beta}}\pa_t+{\frac{\beta}{z}}\pa_\xi\right],\ \ \
e_r= \frac{z}{L}\pa_z,\ \ \
e_i= \frac{z}{L}\pa_i
\eeq
The Levi-Civita connection has non-zero components
\beqn
{\omega^0}_r=\frac{1}{L}\left( e^v-2e^0\right) ,\ \ \
{\omega^0}_v= -\frac{1}{L} e^r,\ \ \
{\omega^v}_r=- \frac{1}{L} e^0, \ \ \
{\omega^i}_r=-\frac1L e^i
\eeqn
Correspondingly, the non-zero components of the Christoffel symbols are
\beqn
\Gamma^z_{zz}=-\frac{1}{z}=\Gamma^\xi_{z\xi}=\Gamma^t_{zt}\\
\Gamma^i_{zj}=-\frac{1}{z}\delta^i_j,\ \ \ \ \Gamma^z_{ij}=\frac{1}{z}\delta_{ij}\\
\Gamma^z_{t\xi}=\frac{1}{z},\ \ \  \Gamma^z_{tt}=-2\frac{\beta^2}{z^3},\ \ \ \Gamma^\xi_{zt}=\frac{\beta^2}{z^3}
\eeqn

\subsection{Dirac Operator}

The spin connection in the spinor representation is obtained by writing
$
{\omega^a}_b=\omega^A {(T^A)^a}_b
$
and replacing the generators by those in the spinor representation. Since the local group is $SO(d+2,1)$, the index $A$ can be thought of as an antisymmetric pair of vector indices. We will use a basis
for the Clifford algebra
\beq
\{ \gamma^a,\gamma^b\}=2\eta^{ab}
\eeq
where $a,b,...=r,0,v,i$. It is always possible to take a basis in which $\gamma^v\gamma^0$ and $\gamma^r$ are {\it Hermitian}.\footnote{In particular, to be definite, we will take $\gamma^0$ to be anti-Hermitian and $\gamma^v$ to be Hermitian.}
For example, a convenient basis is of the form
\beq
\gamma^0=-i\sigma_2\otimes 1,\ \ \
\gamma^v=-\sigma_1\otimes 1,\ \ \
\gamma^i=\sigma_3\otimes \tau^i,\ \ \
\gamma^r=\sigma_3\otimes \tau^r
\eeq
where $\tau^i$ are a representation of $C\ell (d)$.
We have ${(T^{[ab]})^\alpha}_\beta =-\frac{i}{4} {([\gamma^a,\gamma^b])^\alpha}_\beta$. 
Thus, the spin connection takes the form
\beq
{\omega^\alpha}_\beta \sim{\omega^a}_b {([\gamma_a,\gamma^b])^\alpha}_\beta.
\eeq
The $\gamma$'s are numerical matrices and their indices are those of the local frame, raised and lowered with $\eta$.
The Dirac operator in general may be written
\beq
\Dsl=\gamma^c \left(e_c +\frac{1}{4} e_c^\mu {{\omega_\mu}^a}_b (\gamma_a\gamma^b)\right).
\eeq
It will be convenient to split off the radial part
\beqn
\Dsl=z\left[\gamma^r\pa_z+\gamma^i\pa_i\right]+\frac{z^{2\donu{\nu}{}}}{\beta^{\donu{2\nu-1}{}}}(\gamma^v-\gamma^0)\pa_t
+\beta\gamma^v\pa_\xi
-\frac{1}{2}\gamma^r \left[ (d+2\donu{\nu}{})1-\donu{\nu}{}\gamma_v\gamma_0\right].
\eeqn
The problem can be organized by defining projection operators
\beq
P_\pm=\frac{1\pm\gamma^r}{2},\ \ \ \ Q_\pm =\frac{1\pm\gamma^v\gamma^0}{2}
\eeq
These commute with each other, so we can simultaneously diagonalize $\gamma^r$ and $\gamma^v\gamma^0$. We note that these projection operators appear here naturally because {\it they commute with $spin(d)\subset spin(d+2,1)$}. Thus, we can be sure that full $spin(d)$ representations occur for each of the four projection sectors.

Noting then that we can rewrite
\beqn
\gamma^v-\gamma^0=-2Q_+\gamma^0=-2\gamma^0Q_-,\ \ \ \ \gamma^v=(Q_--Q_+)\gamma^0=\gamma^0(Q_+-Q_-)
\eeqn
the Dirac operator becomes
\beq
\Dsl=z\left[(P_+-P_-)\pa_z+\gamma^i\pa_i\right]-2\frac{z^{2\donu{\nu}{}}}{\beta^{\donu{2\nu-1}{}}}\gamma^0Q_-\pa_t
+\beta\gamma^0(Q_+-Q_-)\pa_\xi
-\frac{1}{2}(P_+-P_-) \left[ (d+2\donu{\nu}{})1+\donu{\nu}{}(Q_+-Q_-)\right]
\eeq
Before proceeding with the solution of the Dirac equation, we will consider some details of the bulk representation of the Schr\"odinger algebra that will be important in interpreting the solutions.

\subsection{The Schr\"odinger Algebra}

In a local frame in the bulk, a field will carry a representation of $spin(d+2,1)$. Globally the geometry has isometries given by the Killing vectors
\beqn
N&=&\pa_\xi\\
D&=&2t\pa_t+\vec x\cdot\vec\pa+z\pa_z\\
H&=&\pa_t \\
C&=&t^2\pa_t+t\vec x\cdot\vec\pa-\frac12 \vec x^2\pa_\xi+tz\pa_z-\frac12z^2\pa_\xi\\
M_{ij} &=& x_i\pa_j-x_j\pa_i\\
\vec K&=& -t\vec\pa+\vec x\pa_\xi\\
\vec P&=&\vec \pa
\eeqn
where we have written explicitly the bulk representation on functions. We note that $M_{ij}$ generate $spin(d)$ and that $\{D,H,C\}$ together generate $sl(2,\RR)$. Since $sl(2,\RR)\sim so(2,1)$, we recognize $spin(d)\oplus sl(2,\RR) \subset spin(d+2,1)$. Thus the representation theory can be understood quite simply in terms of highest weight modules. It is traditional to take $D$ diagonal with eigenvalues referred to generically as $\Delta$, and since $N$ is central, it can be diagonalized as well, and in fact there are super-selection sectors labelled by the eigenvalue of $N$
\beq\label{eq:psiequiv}
\Psi(z,t,\xi,\vec x)=e^{in\xi}\Psi(z,t,\vec x).
\eeq
We refer to such functions as being equivariant.\footnote{According to \cite{Duval:2008jg}, the bulk spacetime should be thought of as the total space of a fibre bundle over non-relativistic space-time, with $\xi$ the fibre coordinate. In this sense, eq. (\ref{eq:psiequiv}) is interpreted as meaning that the field is a section of the associated bundle of charge $n$. See also Refs.\cite{Duval:1995fa,Duval:1990hj}.}
Highest weight states of fixed $n,\Delta$ correspond directly to quasi-primary operators $\Psi(t=0,\vec x=0)$ in the boundary theory (at $z=0$).\footnote{Here, by boundary we will simply mean the limit $z\to 0$. Further discussion may be found in a  later section.}

However, for a non-trivial representation, we expect that the generators are modified accordingly, and this structure is not complete. For example, we know (and we will verify below) that the $spin(d)$ generator should be replaced by
\beq\label{eq:Mspin}
M_{ij}=x_i\pa_j-x_j\pa_i+\Sigma_{ij}
\eeq
when acting on spinor fields.
To see how this comes about, we can consider doing Schr\"odinger transformations on the bulk spinor field. The spinor will transform by the Lie derivative\cite{Witten:1983ux}
\beq
\Psi\to {\cal L}_\xi\Psi = \nabla_\xi\Psi+\frac{1}{8}\langle \nabla_{e_a}\xi,e_b\rangle [\gamma^a,\gamma^b]\Psi
\eeq
In this way, ${\cal L}_\xi$ commutes with the Dirac operator, at least as long as $\xi$ is a Killing vector.

Let us warm up by evaluating the Lie derivative for the $spin(d)$ Killing vector
\beq
\xi_M=2x_j \Theta^{ji}\pa_i
\eeq
where $\Theta$ is antisymmetric. In this case, we compute
\beqn
\nabla_{\xi_M}\Psi&=& 2x_j\Theta^{ji}\pa_i\Psi -\frac{1}{z} x_j\Theta^{ji}\gamma_i\gamma^r\Psi\\
\langle \nabla_{e_a}\xi_M,e_b\rangle [\gamma^a,\gamma^b]\Psi&=&4\Theta_{ji}\left[ \gamma^j\gamma^i+2\frac{1}{z}x^j\gamma^i\gamma^r\right]\Psi
\eeqn
 This result does indeed correspond to (\ref{eq:Mspin}).
 Now for $\xi_K$, we have
 \beq
 \xi_K=-t\vec K\cdot\vec\pa+(\vec x\cdot\vec K)\pa_\xi
 \eeq
 and we find
 \beq
{\cal L}_K\Psi = (\vec x\cdot\vec K)\pa_\xi\Psi-t\vec K\cdot\vec\pa\Psi +\frac{1}{2\beta}z(\gamma^0-\gamma^v)K_i\gamma^i\Psi
\eeq
By similar computations, we conclude that the representation on spinors is
\beqn
M_{ij}&=&x_i\pa_j-x_j\pa_i+\frac{1}{4}[\gamma_i,\gamma_j]\equiv x_i\pa_j-x_j\pa_i+i\Sigma_{ij}\label{eq:actionM}\\
K_i&=&-t\pa_i+x_i\pa_\xi +\frac{1}{2}\frac{z}{\beta}(\gamma^0-\gamma^v)\gamma_i\equiv-t\pa_i+x_i\pa_\xi+\kappa_i\label{eq:actionK}\\
C&=&tz\pa_z+t\vec x\cdot\vec\pa+t^2\pa_t-\frac12 (\vec x^2+z^2)\pa_\xi+\hat c\label{eq:actionC}
\eeqn
where $\hat c=+\frac{1}{2}\frac{z}{\beta}(\xsl+z\gamma^r)(\gamma^0-\gamma^v)$. Notice that both $\kappa_i$ and $\hat c$ are nilpotent.
$D,H,P_i$ are unmodified. The highest weight states of a Schr\"odinger multiplet are annihilated by both $C$ and $K_i$, and so we see here that these conditions apparently do not act diagonally on $spin(d)$ components of the bulk Dirac spinor, but mix them at a generic point in the bulk. However, as we will see, the on-shell Dirac spinor is constructed from chiral spinors that are $Q$-chiral and as we show in detail in the appendix, $\hat c$ and $\kappa$ are such that $K_i$ and $C$ act diagonally on these.

\section{Solutions of the Dirac Equation}

Now we write spinors as linear combinations of doubly chiral spinors
\newcommand\eps{\varepsilon}
\beq
\Psi(z,t,\xi,\vec x)=e^{in\xi}\sum_{\eps_r,\eps_\ell}\int \frac{d\omega}{2\pi}\int \frac{d^dk}{(2\pi)^d}e^{i\vec k\cdot\vec x}e^{-i\omega t}\psi_{n,\omega,\vec k}^{\eps_r,\eps_l}(z)
\eeq
where
\beqn
\gamma^r\psi_{n,\omega,\vec k}^{\eps_r,\eps_l}(z)=\eps_r \psi_{n,\omega,\vec k}^{\eps_r,\eps_l}(z)\\
\gamma^v\gamma^0\psi_{n,\omega,\vec k}^{\eps_r,\eps_l}(z)=\eps_l\psi_{n,\omega,\vec k}^{\eps_r,\eps_l}(z)
\eeqn
and $\eps_r,\eps_l=\pm1$. By projecting the Dirac equation with $P_\pm$ and $Q_\pm$, we find four equations (we drop the subscripts on $\psi$ for brevity)
\beqn
\left(z\pa_z-\frac{1}{2}(d+3)+m_0\right)\gamma^0\psi^{-,+}+iz\kslsl\gamma^0\psi^{+,+}+\frac{i}{\beta}(2\omega z^2-n\beta^2)\psi^{+,-}=0\label{eqmp}\\
\left(z\pa_z-\frac{1}{2}(d+1)-m_0\right)\psi^{+,-}+iz\kslsl\psi^{-,-}+in\beta\gamma^0\psi^{-,+}=0\label{eqpm}\\
\left(z\pa_z-\frac{1}{2}(d+3)-m_0\right)\gamma^0\psi^{+,+}-iz\kslsl\gamma^0\psi^{-,+}-\frac{i}{\beta}(2\omega z^2-n\beta^2)\psi^{-,-}=0\label{eqpp}\\
\left(z\pa_z-\frac{1}{2}(d+1)+m_0\right)\psi^{-,-}-iz\kslsl\psi^{+,-}-in\beta\gamma^0\psi^{+,+}=0\label{eqmm}
\eeqn
Note that we have used the notation $\kslsl$ to emphasize that this is the quantity in the trivial metric. The invariant is $\ksl = \gamma^a e_a^i k_i$, and in the Schr\"odinger metric, this evaluates to $\ksl=\frac{z}{L}\kslsl$. This accounts for  the single powers of $z$ accompanying $\kslsl$; this will be of additional importance later in the context of boundary renormalization.

It is not difficult to disentangle the Dirac equations, and we find in particular that
\beq
\left[ z^2\pa_z^2-(d+1)z\pa_z+\left(\frac{d}{2}+1\right)^2-\mu_{\eps_r}^2-q^2z^2\right]\psi^{{\eps_r},-}=0
\eeq
where
\beq
\mu_{\eps_r}=\sqrt{\left(\frac12-{\eps_r}m_0\right)^2+n^2\beta^2}
\eeq
and
\beq
q^2=\vec k^2-2n\omega
\eeq
In this paper, we consider only the case $n\neq 0$.
We note that $q^2$ appears naturally, as it is the Fourier transform of the Galilean-invariant Schr\"odinger operator ${\cal S}=i\pa_t-\frac{1}{2n}\vec\nabla^2$. In particular, we note that when acting on equivariants
\beq
[K_i,{\cal S}]=0
\eeq
This fact will play a central role later in our discussion of renormalizability.

Thus we have \beq \psi^{{\eps_r},-}(\vec k,\omega,z)= z^{1+d/2}
K_{\mu_{\eps_r}}(qz){\bf u}_{\eps_r}(\vec k,\omega) \eeq where ${\bf
u}_\pm$ are independent doubly chiral $spin(d)$ spinors that satisfy
\beq Q_+{\bf u}_\pm=0, \ \ \ \ \ \gamma^r{\bf u}_\pm = \pm {\bf
u}_\pm \eeq

Since we are interested in Euclidean correlator, we have dropped the
solution proportional to $I_{\mu_{\eps_r}}(qz)$ by requiring
regularity at large $z$. Substituting these solutions back into the
Dirac equations leads algebraically to the other components of the
Dirac spinor \beqn \psi^{\pm,+}
&=&\pm\frac{i}{n\beta}z^{1+d/2}\gamma^0\left[
\left(qzK_{\mu_\mp}'(qz)+\left(\frac12\pm
m_0\right)K_{\mu_\mp}(qz)\right){\bf u}_\mp(\vec k,\omega)\mp iz
K_{\mu_\pm}(qz)\kslsl{\bf u}_\pm(\vec k,\omega)\right]\nonumber
\eeqn The general on-shell field then is \beqn \Psi=
-z^{1+d/2}\sum_{\eps_r}\left[\frac{i\eps_r}{n\beta}\left(\frac12-\eps_rm_0-i\eps_rn\beta\gamma^0\right)K_{\mu_{\eps_r}}(q
z)+\frac{i\eps_r}{n\beta}qzK'_{\mu_{\eps_r}}(q
z)+\frac{z}{n\beta}K_{\mu_{\eps_r}}(q z)\kslsl\right]\gamma^0{\bf
u}_{\eps_r} \eeqn

From this general solution, we see that the leading terms at the
boundary are \beq \Psi\sim \sum_{\eps_r}z^{\Delta^-_{\eps_r}}
\frac{\Gamma(\mu_{\eps_r})}{2}\left(\frac{q}{2}\right)^{-\mu_{\eps_r}}X_{-(0)}^{\eps_r}{\bf
u}_{\eps_r} +\ldots \eeq where $X^{\eps_r}_{-(0)}\equiv 1 -
\frac{i\eps_r}{n\beta}(\frac{1}{2} - \eps_r m_0 -
\mu_{\eps_r})\gamma^0 $
 and where \beq \Delta_{\eps_r}^\pm =1+
\frac{d}{2}\pm\mu_{\eps_r}, \eeq
We make several comments:
\begin{itemize}
\item For generic bulk mass, the two dimensions $\Delta_{\eps_r}^-$ are irrationally related. Thus
${\bf u}_{+}$ and ${\bf u}_{-}$  must be taken as independent sources. The only counter-example (for $n\neq 0$) is if the bulk mass vanishes, in which case $\mu_+=\mu_-$ and $\Delta^-_+=\Delta^-_-\equiv \Delta^-$. Thus the massless case is special, in that the eigenvalues of $D$ are degenerate. One can show however that in either case, ${\bf u}_\pm$ transform separately under the Schr\"odinger algebra.
\item More precisely, the coefficient of the leading singularity is of the form $X_{-(0)}^{\eps_r}{\bf u}_{\eps_r}$ and is not chiral. We note though that the $X_{-(0)}^{\eps_r}$ are {\it constant} matrices and can be absorbed (by a basis change) into the definition of boundary operators.
\item The dependence on $\kslsl$ is subleading in the near-boundary ($z\to 0$) expansion, and is associated with the {\it odd} powers of $z$.
\item The general solution is obtained by specifying two $spin(d)$ spinors ${\bf u}_\pm$ that are $Q$-chiral  ($Q_+ {\bf u}_\pm=0$). Thus the Dirac equation has eliminated half of the degrees of freedom, as expected.
\item The Schr\"odinger covariance of this expression is not manifest as written. To see this, consider the symplectic 1-form
\beq
\alpha = -i dx^\mu\otimes \pa_\mu
\eeq
which when acting on plane waves  gives
\beq
\alpha = -idz\otimes\pa_z + (nd\xi -\omega dt+\vec k\cdot d\vec x)
\eeq
The scalar Lagrangian, for example, can be written
\beq
S_{scalar}\sim \int \sqrt{-g} \langle \alpha(\phi),\alpha(\phi)\rangle = -\int\sqrt{-g}\phi^\dagger \Delta\phi
\eeq
For fermions, we replace the exterior algebra by the Clifford algebra, and hence we obtain
\beq
\alpha\to -i\frac{z}{L}\gamma^r\pa_z+\frac{n\beta}{L}\gamma^v+\omega \frac{z^2}{L\beta}(\gamma^0-\gamma^v)+\frac{z}{L}\kslsl
\eeq
This of course is the quantity appearing in the Dirac operator, apart from the spin connection terms. What we learn from this is that the $\kslsl$ term should really be grouped with
\beq\label{eq:qsl}
\qsl=z\kslsl+n\beta\gamma^v+2\omega \frac{z^2}{\beta}\gamma^0Q_-
\eeq
In particular, this combination is Galilean invariant, $[K_i,\qsl]=0$, and also $[D,\qsl]=0$. (Later, it will play a central role in the renormalizability of the theory.) We also note that
\beq
\qsl^2=z^2{\cal S}+n^2\beta^2
\eeq
which is the scalar invariant noted above, where ${\cal S}=q^2=\vec k^2-2n\omega$.

Furthermore, when acting on the chiral spinors $\gamma^0{\bf u}_\pm$, this simplifies to
\beq
\qsl\gamma^0{\bf u}_\pm=n\beta \left(\gamma^0+\frac{z\kslsl}{n\beta}\right)\gamma^0{\bf u}_\pm
\eeq
and thus the on-shell spinor can be rewritten
\beqn\label{eq:onshellfield}
\Psi=
-\frac{i}{n\beta}z^{1+d/2}\sum_{\eps_r}\left[\eps_rqzK'_{\mu_{\eps_r}}(q z)-(m_0-\eps_r/2)K_{\mu_{\eps_r}}(qz)-i K_{\mu_{\eps_r}}(q z)\qsl\right]\gamma^0{\bf u}_{\eps_r}
\eeqn
We note that ${\bf v}_{\eps_r}\equiv\gamma^0{\bf u}_{\eps_r}$ are chiral, with $Q_-{\bf v}_{\eps_r}=0$.
 We note though that the individual pieces of $\qsl$ come in with different powers of $z$, and so going to the boundary is somewhat subtle. What we must do is understand how the generators of the Schr\"odinger algebra act on the terms in the expansion of the field. This is explained in detail in the Appendix.

 \end{itemize}

\section{Variational Principle and Boundary Renormalization}

As in the relativistic case
\cite{Henningson:1998cd,Mueck:1998iz,Henneaux:1998ch,Giecold:2009tt}, we need to
add a boundary term to the Dirac action, as the bulk part of the Dirac action vanishes on-shell. This boundary term serves to give
the proper Dirichlet boundary condition and simultaneously make the
on-shell action finite. Since in the variational principle the field
variations are off-shell, first of all we have to state clearly what
we mean by an off-shell spinor. The on-shell solution
(\ref{eq:onshellfield}) suggests that an off-shell spinor should
have the same $z$ expansion, except that the coefficients are in
general full unconstrained Dirac spinors. This off-shell spinor obviously carries
a Schrodinger representation through
(\ref{eq:actionM})-(\ref{eq:actionC}). However, as established in
the Appendix, this representation is reducible. Hence, it is natural that
we define the off-shell spinor to be an irreducible representation
that encompasses all vacuum solutions (\ref{eq:onshellfield}).
According to (\ref{eq:offshellfield-app}), it takes the form \beqn
\Psi= \sum_{\eps_r}\left[ z^{\Delta^-_{-\eps_r}}\sum_{k=0}^\infty
z^{2k}\left(\rho^{\eps_r}_{(2k)} + \qsl\rho_{(2k+1)}^{\eps_r}\right)
+z^{\Delta^+_{-\eps_r}}\sum_{k=0}^\infty
z^{2k}\left(\chi^{\eps_r}_{(2k)} + \qsl\chi_{(2k+1)}^{\eps_r}\right)
\right]\label{eq:offshellfield}\eeqn where $\qsl$ is given by
(\ref{eq:qsl}) and
\beq \gamma^r\rho_{(m)}^{\eps_r}=\eps_r\rho_{(m)}^{\eps_r},\ \ \ \
Q_-\rho_{(m)}^{\eps_r}=0 \eeq similarly for $\chi$.

By comparison to the on-shell solution, one can deduce the on-shell
relationship of $\rho_{(m)}^{\eps_r}$ and $\chi_{(m)}^{\eps_r}$ to
the independent quantities $\rho_{(0)}^{\eps_r}\sim {\bf
u}_{\eps_r}$ \beqn
\rho_{(2k)}^{\eps_r}&=& -ix_{-(2k)}^{\eps_r}\rho^{\eps_r}_{(2k+1)} = \frac{x^{\eps_r}_{-(2k)}}{x^{\eps_r}_{-(0)}}\frac{\Gamma(1-\mu_{-\eps_r})}{k!\Gamma(k+1-\mu_{-\eps_r})}\left(\frac{q}{2}\right)^{2k}\rho_{(0)}^{\eps_r}\\
\chi_{(2k)}^{\eps_r}&=&
-ix_{+(2k)}^{\eps_r}\chi^{\eps_r}_{(2k+1)}=-\frac{x^{\eps_r}_{+(2k)}}{x^{\eps_r}_{-(0)}}\frac{\Gamma(1-\mu_{-\eps_r})}{k!\Gamma(k+1+\mu_{-\eps_r})}\left(\frac{q}{2}\right)^{2k
+ 2\mu_{-\eps_r}}\rho_{(0)}^{\eps_r}, \eeqn where
$x^{\eps_r}_{\pm(2k)} = \eps_r(\frac{1}{2} + \eps_r m_0 \pm
\mu_{-\eps_r} + 2k).$

As is well known, the bulk part of the Dirac action evaluates to
zero on the equations of motion. The on-shell action is determined
entirely by the boundary term. The renormalized Dirac action must be
of the form \beq\label{eq:onshellaction} S_{Lor}=\int_M
d^{d+3}x\sqrt{-g}\ \overline\Psi i(\Dsl-m_0)\Psi
+\frac{1}{L^2}\int_{\pa M}dt\ d\xi\ d^{d}x\sqrt{\gamma}\
e_r^z\overline\Psi T\Psi \eeq for some matrix $T$, which must
respect the symmetries of the boundary theory. Given that \beq
\Psi(z,t,\xi,\vec x)=e^{in\xi}\Psi(z,t,\vec x) \eeq the action
reduces to \beq R\int d^{d+1}x\int dz\sqrt{-g}\ \overline\Psi
i(\Dsl_n-m_0)\Psi +R\int dt\ d^{d}x\sqrt{\gamma}\ z^{-d-2}\
\overline\Psi T\Psi \eeq where $\gamma$ denotes the spatial induced
$d$-metric of the boundary (which in our case is flat). We absorb
this overall factor of $R$ into the normalization. We vary the
action subject to the vanishing of the variation of the source.
Given our choice of action, we find \beq\label{eq:varyaction} \delta
S=\int  \bar\Psi(i\gamma^r+T)\delta\Psi+\int \delta\bar\Psi T\Psi
\eeq A proper variational principle is obtained by requiring that
terms involving $\delta\chi$ not appear in this expression. This
will force the variational principle to give the correct Dirichlet
condition $\delta\rho=0$ on the boundary. In addition to this
requirement, the resulting on-shell action must be made finite by
the addition of suitable boundary counter-terms. As is customary, we
will use minimal subtraction. By suitable, we mean any term that
respects the symmetry of the regulated boundary theory. In the case of AdS,
the boundary  counterterms are Poincar\'e invariant, which is the
symmetry respected by the regulator. In our case, we expect the
counterterms to be Galilean invariant. Since these counterterms are
all written in terms of the boundary values of bulk fields, upon
which the Schr\"odinger transformations act in the prescribed way,
it is appropriate to write the boundary counterterms as boundary
values of bulk-invariant terms (that is, using the bulk metric for
contractions and the bulk realization of the symmetry generators).

Requiring the boundary term to be Galilean invariant, $T$ has to be
written in terms of operators that commute with the Galilean
generators, in particular the $K_i$. Careful consideration of this
problem reveals that such invariants may be constructed out of
$P_{\eps_r}$, $\gamma^0Q_-$ and $\qsl$ and thus the most general
boundary term can be written as (here
$L=\qsl^2/n^2\beta^2$)\footnote{the similar expression for AdS would
be $\sum_{\eps_r} (a_{\eps_r}(q^2) + \ksl\hspace{2pt}
b_{\eps_r}(q^2))P_{\eps_r}$, where $P_{\eps_r}$ is the projector
along the radial direction.} \beq T=\sum_{\eps_r} \left[
a_{\eps_r}(L)+b_{\eps_r}(L)\frac{i\qsl}{n\beta}+\left[
c_{\eps_r}(L)+d_{\eps_r}(L)\frac{i\qsl}{n\beta}\right]\gamma^0Q_-\right]P_{\eps_r}
\eeq where $a_{\eps_r}(L),...$ are functions to be determined.
Although we have written the coefficient functions as functions of
$\qsl^2\sim L$ for notational brevity, since $N$ is central and the
fields are equivariant, this could just as well be replaced by
$z^2{\cal S}=z^2q^2$.

This form for $T$ and the field written in the form
(\ref{eq:onshellfield}) is most convenient to discuss the
renormalizability of the theory -- it organizes the counterterms in
an invariant fashion. What is more complicated here, compared to the
relativistic case, is that this organization is not homogeneous in
powers of $z$. It is easy to see in this form however how the
renormalization will work -- since $\qsl^2\sim L$, we can regard $T$
as an expansion in powers of $\qsl$ (rather than $z$). At any given
order, canceling divergences will correspond to conditions on the
Taylor coefficients of the functions $a_{\eps_r(L)},...$ around $L
=1$. Depending on the values of various parameters ($m_o$, $n$, ...), we can terminate
the Taylor expansion at some order, as all further contributions to
the action will be zero when the cutoff is removed. It remains then
to demonstrate that the conditions on the Taylor coefficients can be
consistently solved to remove all divergences. We will not construct
a general proof, and in fact will work just at lowest order.
Experience with these manipulations suggests that no problems will
be encountered at higher orders.

Given the form for $T$, we consider the variational problem; this
will place conditions on the lowest order Taylor coefficients of the
functions in $T$. We write (\ref{eq:varyaction}) in terms of
$\delta\rho_{(m)}^{\eps_r}$ and $\delta\chi_{(m)}^{\eps_r}$ and
their conjugates. Due to the fact that $\mu_- - \mu_+ < 1$ for $n
> 0$, the only terms that possibly contain $\delta\chi_{(m)}^{\eps_r}$ are the
finite term ($z^0$ power), which evaluate to
\beqn
(\rho_{(0)}^{\eps_r})^\dagger\delta\chi_{(0)}^{\eps_r}\left[-ib_{\eps_r}-\frac{in\beta}{x_{-(0)}^{\eps_r}}(a_{\eps_r}+i{\eps_r})\right]
+n\beta(\rho_{(0)}^{\eps_r})^\dagger\delta\chi_{(1)}^{\eps_r}\left[i\eps_r-
(a_{-\eps_r}-id_{-\eps_r})+\frac{n\beta}{x_{-(0)}^{\eps_r}}(b_{-\eps_r}+ic_{-\eps_r})\right]
\eeqn
where $a_{\eps_r}\equiv a_{\eps_r}(1)$, etc. To obtain this
expression, we have used the on-shell relations for the
$\rho_{(m)}^{\eps_r}$. Similarly, the only terms involving
$\delta\chi^\dagger$ are
\beq
-(\delta\chi^{\eps_r}_{(0)})^\dagger\rho_{(0)}^{\eps_r}\left[ib_{\eps_r}+\frac{in\beta}{x_{-(0)}^{\eps_r}}(a_{-\eps_r}-id_{-\eps_r})\right]+
n\beta(\delta\chi^{\eps_r}_{(1)})^\dagger\rho_{(0)}^{\eps_r}\left[a_{\eps_r}-\frac{n\beta}{x_{-(0)}^{\eps_r}}(b_{-\eps_r}+ic_{-\eps_r})\right]
\eeq
The variational principle requires each term to vanish separately, which results in three independent equations
\beqn
b_{\eps_r} &=& -\frac{n\beta}{x^{\eps_r}_{-(0)}}(a_{\eps_r} +
i\eps_r)\label{cond-1}\\
b_{\eps_r} &=& -\frac{n\beta}{x^{\eps_r}_{-(0)}}(a_{-\eps_r} -
id_{-\eps_r})\label{cond-2}\\
a_{\eps_r} &=&
\frac{n\beta}{x^{\eps_r}_{-(0)}}(b_{-\eps_r}+ic_{-\eps_r})\label{cond-3}
\eeqn As argued above, given a specific value of $a_{\eps_r} =
a_{\eps_r}(1),...$ satisfying (\ref{cond-1})-(\ref{cond-3}), the
higher order coefficients in the Taylor expansion can be found
successively by requiring the cancelation of subleading divergences.
There are, however, two things that those higher order coefficients
cannot control, since they involve subleading powers in $z$. They
are the leading divergence, of the form
$\rho^{\eps_r\dagger}_{(0)}\rho^{\eps_r}_{(0)}$\footnote{Other
possible terms such as
$\rho^{\eps_r\dagger}_{(0)}\kslsl\rho^{-\eps_r}_{(0)}$ must be
thought of as a piece of $\bar\rho^{\eps_r}_{(0)}\gamma^0\qsl
\rho^{-\eps_r}_{(0)}$, but as we have discussed, this is not $K_i$
invariant, so will not appear.}, and the finite part of the on-shell
action. Requiring the leading divergence to be zero sets \beq
-ib_{\eps_r}-\frac{in\beta}{x_{-(0)}^{\eps_r}}(a_{-\eps_r}-id_{-\eps_r})
+\frac{in\beta}{x^{\eps_r}_{-(0)}}\left(a_{\eps_r}-\frac{n\beta}{x_{-(0)}^{\eps_r}}(b_{-\eps_r}+ic_{-\eps_r})\right)
= 0, \eeq which is fortunately automatically satisfied from
(\ref{cond-2}) and (\ref{cond-3}). The finite part of the on-shell
action is in fact independent of the values of
$a_{\eps_r}$,... (although the variation of the action is not).
Indeed, a short calculation gives \beqn
\nonumber S_{os}&=&\int \frac{d\omega}{2\pi}\int \frac{d^dk}{(2\pi)^d}\sum_{\eps_r} \frac{2\eps_r\mu_{-\eps_r}}{n\beta}(\rho_{(0)}^{\eps_r})^\dagger  \chi_{(0)}^{\eps_r}\\
&=&-\frac{2\eps_r\mu_{-\eps_r}}{n\beta}\int \frac{d\omega}{2\pi}\int
\frac{d^dk}{(2\pi)^d}\sum_{\eps_r}\frac{x^{\eps_r}_{+(2k)}}{x^{\eps_r}_{-(0)}}\frac{\Gamma(1-\mu_{-\eps_r})}{\Gamma(1+\mu_{-\eps_r})}\left(\frac{q^2}{4}\right)^{\mu_{-\eps_r}}(\rho_{(0)}^{\eps_r})^\dagger
\rho_{(0)}^{\eps_r},\eeqn where we have used the conditions
(\ref{cond-1})-(\ref{cond-3}). Thus, it is scheme independent.

\section{Boundary Operators}

As we have seen, the leading term in the expansion of the field is
proportional to $X_{-(0)}^{\eps_r}\rho_{(0)}^{\eps_r}$, where
$X_{-(0)}^{\eps_r}$ is a constant matrix and $\rho_{(0)}^{\eps_r}$
is a (doubly) chiral spinor field. It is convenient to take a basis
of boundary quasi-primary operators such that $\rho_{(0)}^{\eps_r}$
act as the sources for operators of charge $n$ and dimension
$\Delta^+_{\eps_r}$ \beq\label{eq:opcouple} \int dt \int
d^dx\sqrt{\gamma}\ \left[ (\rho_{(0)}^{\eps_r})^\dagger(t,\vec x)
{\cal O}_{n,\eps_r}(t,\vec x)+h.c.\right] \eeq This is possible
because the $X_{-(0)}^{\eps_r}$ are constant matrices. This coupling
preserves the Schr\"odinger invariance at the boundary, obtained
from the bulk transformations, for example \beqn
v^iK_i:&\Psi'(t',\vec x',z')=e^{in(\vec v\cdot\vec x+i\vec v^2 t/2)}(1+v\cdot\kappa)\Psi(t,\vec x,z)\label{eq:transfieldK}\\
cC:&\Psi'(t',\vec x',z')=e^{-\frac{inc}{2}\frac{\vec x^2+z^2}{1
+ct}}(1+c\hat c)\Psi(t,\vec x,z)\label{eq:transfieldC}, \eeqn at
$z=0$. Under, say, a finite $C$ transformation, the coupling
(\ref{eq:opcouple}) transforms as \beq \int dt \int
d^dx\sqrt{\gamma}\ (\rho_{(0)}^{\eps_r})^\dagger(t,\vec x) {\cal
O}_{n,\eps_r}(t,\vec x) \to \int \frac{dt
d^dx}{(1+ct)^{d+2}}(1+ct)^{\Delta_{-\eps_r}^++\Delta_{-\eps_r}^-}
(\rho_{(0)}^{\eps_r})^\dagger(t,\vec x) {\cal O}_{n,\eps_r}(t,\vec
x) \eeq given the appropriate transformation of boundary
quasi-primary operators (see for example Ref. \cite{Henkel:2003pu}).
Thus the coupling is invariant, as
$\Delta_{\eps_r}^++\Delta_{\eps_r}^-=d+2$.

Given the form of the on-shell action, we then read off the
two-point Euclidean correlator of quasi-primary operators
\beq
\langle ({\cal O}_{n,\eps_r})^\dagger (t,\vec x){\cal
O}_{n',\eps'_r} (t',\vec x')\rangle=
-\delta_{\eps_r,\eps'_r}\delta_{n,n'} \int \frac{d\omega}{2\pi}
\frac{d^dk}{(2\pi)^d} e^{-i\omega(t'-t)}e^{i\vec k\cdot(\vec x'-\vec
x)}\frac{2\eps_r\mu_{-\eps_r}}{n\beta}\frac{x^{\eps_r}_{+(2k)}}{x^{\eps_r}_{-(0)}}\frac{\Gamma(1-\mu_{-\eps_r})}{\Gamma(1+\mu_{-\eps_r})}\left(\frac{q^2}{4}\right)^{\mu_{-\eps_r}}
\eeq
By scaling, it is easy to see that this behaves as
\[(t'-t)^{-\Delta_{-\eps_r}^+} f\left(\frac{(\vec x'-\vec
x)^2}{(t'-t)}\right),\] and in fact this is just proportional to the
scalar propagator. We note that this correlator
preserves chirality, and in particular no $\gamma$-matrix structure
is present. This is expected of a non-relativistic theory, as there
is no essential difference between boson and fermion fields.

This is not to say that other correlation functions do not have more interesting structure. The subleading terms in the asymptotic expansion of the field are sources for descendant operators. Given the form of the generators in the bulk  (\ref{eq:actionK},\ref{eq:actionC}), we see that Schr\"odinger transformations mix the descendant fields in an interesting way. 
The correlation functions of dual operators will of course display a similar structure, and thus these correlation functions can have non-trivial $\gamma$-matrix structure.

\section{Conclusion}

We have investigated carefully the Dirac fermion problem on the spacetime of Schr\"odinger isometry, which is dual to the vacuum configuration of non-relativistic conformal field theories in $d$ spatial dimensions. The structure of the system is rather intricate, but a sensible Dirichlet problem exists and the boundary theory is renormalizable.

Although the bulk geometry contains a compact null direction (coordinatized by $\xi$), the metric is of the Bargman type and the usual holographic prescriptions go through more or less unmodified for equivariant operators, with care taken in interpreting the boundary action. The bulk field sources operators of a highest weight module of the Schr\"odinger algebra and the correct structure of two-point correlation functions is obtained.
It would be interesting to extend these computations to higher point functions and to finite density.

We should mention that we are aware of one paper concerning fermions in the Schr\"odinger geometry \cite{Akhavan:2009ns}. We have not been able to understand the details of the computations in this paper or their results.

\section{Acknowledgments}

It is a pleasure to thank Allan Adams, Juan Jottar, Djordje Minic, Leo Pando-Zayas and Tassos Petkou for conversations on related matters and especially Mohammad Edalati for assistance at an early stage of this project.
RGL thanks the Aspen Center for Physics and the organizers and participants of the workshop ``String Duals of Finite Temperature and Low-Dimensional Systems" 
where some of this research was carried out. Work supported in part by the US Department of Energy, under contract DE-FG02-91ER40709.

\appendix
\section{Appendix}
\renewcommand{\theequation}{A.\arabic{equation}}
\setcounter{equation}{0}
\newcommand{\er}{\eps_r}
\newcommand{\el}{\eps_\ell}
\newcommand{\erp}{\eps'_r}
\newcommand{\elp}{\eps'_\ell}
\newcommand{\mer}{\bar\eps_r}
\newcommand{\mel}{\bar\eps_\ell}


As noted in the text, the realization in the bulk of $K_i$ and $C$ acting on fermions is
\begin{align}
K_i =& -t\pa_i + x_i \pa_\xi + \frac{z}{\beta}\gamma^0\gamma_i Q_-\\
=& K^{(0)}_i + \frac{z}{\beta}\gamma^0\gamma_i Q_-\\
C =& t(z\pa_z + \vec x \vec\pa + t\pa_t) - \frac{1}{2}(\vec x^2 +
z^2)\pa_\xi + \frac{z}{\beta}(\xsl + z\gamma^r)\gamma^0 Q_-\\
=& C^{(0)}  - \frac{1}{2} z^2\pa_\xi +  \frac{z}{\beta}(\xsl +
z\gamma^r)\gamma^0 Q_-
\end{align}
\subsection{off-shell transformation}

Motivated by the general solution of the Dirac equation, the
off-shell spinors are assumed to have the near boundary expansion
\begin{align}\label{off-shell psi}
\nonumber \Psi =& z^{\Delta^-_+}\sum_{k=0}^{\infty}
z^{2k}(\Psi^I_{(2k)} + z \Psi^I_{(2k+1)}) +
z^{\Delta^+_+}\sum_{k=0}^{\infty} z^{2k} (\Psi^{II}_{(2k)} +
z \Psi^{II}_{(2k+1)}) \\
&\h{100pt}+ z^{\Delta^-_-}\sum_{k=0}^{\infty} z^{2k}
(\Psi^{III}_{(2k)} + z \Psi^{III}_{(2k+1)})+
z^{\Delta^+_-}\sum_{k=0}^{\infty} z^{2k}(\Psi^{IV}_{(2k)} + z
\Psi^{IV}_{(2k+1)})
\end{align}
containing four power series of $z$, in which the $\Psi_{(m)}$'s are in
general full Dirac spinors. For generic values of $d$, $m_0$ and $n$
these series do not talk to each other under Schr\"odinger
transformations. Each of them form a separate representation of the
Schr\"odinger group. As our purpose is to work out the transformation
laws, it is sufficient to focus on just one of them, say the one with
$\Delta^-_+$. Results for the other series are inferred immediately.

Let's re-parametrize the first series in terms of $P$ and $Q$ chiral
spinors as follows
\begin{equation}\label{PsiI}
\Psi^I = \sum_{k,\erp\elp}z^{\Delta^-_+ +
2k}\Big{(}\rho^{\erp\elp}_{(2k)} +
\qsl\h{3pt}\rho^{\erp\elp}_{(2k+1)}\Big{)},
\end{equation}
where $\gamma^r \rho^{\er\el}_{(m)} = \er \rho^{\er\el}_{(m)}$ and
$\gamma^v \gamma^0 \rho^{\er\el}_{(m)} = \el \rho^{\er\el}_{(m)}$.

Our task is to find the transformation laws of $\rho^{\er\el}_{(m)}$
under the Schr\"odinger algebra, restricted to the non-trivial
isometries $K_i$ and $C$. We will then argue that it is possible to consistently
reduce the representation by setting half of the fields to zero,
leaving the $\rho^{\er +}_{(m)}$ untouched. At this point, the
remaining fields $\rho^{++}_{(m)}$ and $\rho^{-+}_{(m)}$ transform
independently, so one of them can be further set to zero under the
criteria that the leftover representation should include all
on-shell solutions. In particular, a suitable irreducible representation corresponds to keeping
$\rho^{-+}_{(m)}$ for $\Psi^I$, $\Psi^{II}$ and $\rho^{++}_{(m)}$
for $\Psi^{III}$, $\Psi^{IV}$.

Transformations of $\rho^{\er\el}_{(m)}$ are straightforwardly found
by acting with $K_i$ and $C$ on (\ref{PsiI}) and reading off the coefficients
of different powers of $z$. Projecting further by
 $P_{\eps_r}Q_{\eps_\ell}$ we get
\begin{align}
\ksl\delta_{K_i}\rho^{\mer\el}_{(2k+1)} =& \ksl
K_i^{(0)}\rho^{\mer\el}_{(2k+1)} + \frac{1}{\beta}\gamma^0\gamma_i
\rho^{\er -}_{(2k)} \delta_{\el,+} - n \gamma_i \rho^{\mer -
}_{(2k+1)}
\delta_{\el,-}\label{K-odd}\\
\ksl\delta_{C}\rho^{\mer\el}_{(2k+1)} =&
C^{(0)}\ksl\h{2pt}\rho^{\mer\el}_{(2k+1)} +
\frac{\xsl}{\beta}\gamma^0 \rho^{\er -}_{(2k)} \delta_{\el,+} -
\frac{in}{2}\ksl\h{3pt}\rho^{\mer\el}_{(2k-1)} -
n\xsl\h{3pt}\rho^{\mer +}_{(2k+1)} \delta_{\el,+} +
\frac{1}{\beta}\gamma^r\gamma^0 \ksl\h{2pt}\rho^{\er
-}_{(2k-1)}\delta_{\el,+}\label{C-odd}
\end{align}
and
\begin{align}
\nonumber \delta_{K_i}\rho^{\er\el}_{(2k)} + n\beta \gamma^v
\delta_{K_i}\rho^{\mer\mel}_{(2k+1)} + \frac{2\omega}{\beta}\gamma^0
\delta_{K_i}\rho^{\mer -}_{(2k-1)} &\delta_{\el,+}  = K_i^{(0)}
\Big{(}\rho^{\er\el}_{(2k)} + n\beta \gamma^v
\rho^{\mer\mel}_{(2k+1)} + \frac{2\omega}{\beta}\gamma^0 \rho^{\mer
-}_{(2k-1)} \delta_{\el,+} \Big{)}\\
&\h{110pt} - \frac{1}{\beta}\ksl \gamma^0\gamma_i\rho^{\mer
-}_{(2k-1)}\delta_{\el,+}\label{K-even}\\
\nonumber \delta_{C}\rho^{\er\el}_{(2k)} + n\beta \gamma^v
\delta_{C}\rho^{\mer\mel}_{(2k+1)} + \frac{2\omega}{\beta}\gamma^0
\delta_{C}\rho^{\mer -}_{(2k-1)} &\delta_{\el,+}  = C^{(0)}
\Big{(}\rho^{\er\el}_{(2k)} + n\beta \gamma^v
\rho^{\mer\mel}_{(2k+1)} + \frac{2\omega}{\beta}\gamma^0 \rho^{\mer
-}_{(2k-1)} \delta_{\el,+} \Big{)}\\
\nonumber &-\frac{in}{2}\rho^{\er\el}_{(2k-2)} +
\frac{1}{\beta}\gamma^r\gamma^0\rho^{\mer -}_{(2k-2)} \delta_{\el,+}
- \frac{in^2\beta}{2}\gamma^v \rho^{\mer\mel}_{(2k-1)} \\
&- \frac{in\omega}{\beta}\gamma^0 \rho^{\mer
-}_{(2k-3)}\delta_{\el,+} +
\frac{\xsl}{\beta}\gamma^0\ksl\h{3pt}\rho^{\mer
-}_{(2k-1)}\delta_{\el,+} - \er n\rho^{\er
+}_{(2k-1)}\label{C-even}.
\end{align}
Here $\bar\eps_{r,\ell} = - \eps_{r,\ell}$ and
$\rho^{\eps'_r\elp}_{(m)}$, $m < 0$ are defined to be zero. It is
also important to note that the action of $C^{(0)}$ (the part of $C$ acting on functions), due to the term
$z\pa_z$, depends on the dimension of the fields it acts on. From
(\ref{K-odd}) and (\ref{C-odd}), we see that $\rho^{\er -}_{(2k-1)}$
transform into themselves
\begin{align}
\ksl\delta_{K_i}\rho^{\mer -}_{(2k+1)} =& \ksl
K_i^{(0)}\rho^{\mer -}_{(2k+1)} - n \gamma_i \rho^{\mer - }_{(2k+1)} \label{K-odd-}\\
\ksl\delta_{C}\rho^{\mer -}_{(2k+1)} =& C^{(0)}\ksl\h{2pt}\rho^{\mer
- }_{(2k+1)}  - \frac{in}{2}\ksl\h{3pt}\rho^{\mer -}_{(2k-1)}
\label{C-odd-},
\end{align}
hence can be consistently set to zero. With that in mind, the
transformations of $\rho^{\er -}_{(2k)}$ can be deduced from
(\ref{K-odd}), (\ref{C-odd}), (\ref{K-even}) and (\ref{C-even}) to
read
\begin{align}
\ksl\delta_{K_i}\rho^{\er -}_{(2k)} =& \ksl K_i^{(0)}\rho^{\er
-}_{(2k)} - n\gamma_i \rho^{\er -}_{(2k)}\label{K-even-}\\
\ksl\delta_{C}\rho^{\er -}_{(2k)}) =& \ksl C^{(0)}\rho^{\er -
}_{(2k)} - \frac{in}{2}\ksl\h{3pt}\rho^{\er -}_{(2k-2)} +
n\xsl\h{2pt} \rho^{\er -}_{(2k)} \label{C-odd-}.
\end{align}
Again, they only transform into themselves. Thus, we have shown that
the representation can be reduced by setting $\rho^{\er -}_{(m)}  =
0$. The transformations of the remaining fields are
simplified significantly
\begin{align}
\delta_{K_i}\rho^{\er +}_{(2k+1)} =& K_i^{(0)}\rho^{\er
+}_{(2k+1)} \label{K-odd+}\\
\delta_{C}\rho^{\er +}_{(2k+1)} =& C^{(0)}\rho^{\er + }_{(2k+1)}
- \frac{in}{2}\h{3pt}\rho^{\er +}_{(2k-1)} \label{C-odd+}\\
\delta_{K_i}\rho^{\er +}_{(2k)} =& K_i^{(0)}\rho^{\er
+}_{(2k)} \label{K-even+}\\
\delta_{C}\rho^{\er +}_{(2k)} =& C^{(0)}\rho^{\er + }_{(2k)} -
\frac{in}{2}\h{3pt}\rho^{\er +}_{(2k-2)} - \er n \rho^{\er
+}_{(2k-1)} \label{C-even+}
\end{align}
Looking at (\ref{K-odd+})-(\ref{C-even+}), we notice that fields
with opposite $\er$ index do not mix under the transformations as
well. Thus, the representation can be maximally reduced by setting
one of the two $\rho^{\er +}_{(m)}$ to zero. The criteria is
obvious: the leftover representation must include all solutions of
the Dirac equation. Thus, in (\ref{off-shell psi}) it corresponds to
keeping only $\rho^{\mer +}_{(m)}$ in the series with leading order
$z^{\Delta^\pm_{\er}}$. This gives the off-shell spinor \beqn \Psi=
\sum_{\eps_r}\left[ z^{\Delta^-_{-\eps_r}}\sum_{k=0}^\infty
z^{2k}\left(\rho^{\eps_r}_{(2k)} + \qsl\rho_{(2k+1)}^{\eps_r}\right)
+z^{\Delta^+_{-\eps_r}}\sum_{k=0}^\infty
z^{2k}\left(\chi^{\eps_r}_{(2k)} + \qsl\chi_{(2k+1)}^{\eps_r}\right)
\right]\label{eq:offshellfield-app},\eeqn where $\rho^{\eps_r} =
\rho^{\eps_r +}$, etc.

\subsection{Massless limit}

In the massless limit, $\mu_+ = \mu_- = \mu$, $\Delta^\pm_- =
\Delta^\pm_+ = \Delta^\pm$ and the expansion (\ref{off-shell psi})
collapses into just two series. However, for each series all the
analysis carried out  above is still valid. The only
difference is that now to include all solutions of the Dirac
equation in the reduced representation, for each series we must keep
both $\rho^{\er +}_{(m)}$ rather than just one of them. The
transformation laws are the same as (\ref{K-odd+})-(\ref{C-even+}).
Again, fields with opposite $\er$ index do not mix under the
transformations. Each series then contains two irreducible
representations of the Schr\"odinger group, labeled by $\er$.


\providecommand{\href}[2]{#2}\begingroup\raggedright\endgroup

\end{document}